\documentclass[aip,jcp,floatfix,preprint,10pt,a4paper]{revtex4-1}
\usepackage[dvips]{graphicx}
\usepackage{epsfig}

\usepackage{verbatim}

\usepackage{times}

\usepackage{subfigure}

\usepackage{amsmath}

\DeclareMathOperator{\sgn}{sgn}

\begin{document}


\title{Statistical mechanics of Roskilde liquids: configurational adiabats, specific heat contours and density dependence of the scaling exponent}
\date{\today}
\author{Nicholas P. Bailey}
\author{Lasse B{\o}hling}
\author{Arno A. Veldhorst}
\author{Thomas B. Schr{\o}der}
\author{Jeppe C. Dyre}
\affiliation{DNRF Center "Glass and Time", IMFUFA, Dept. of Sciences, Roskilde University, P. O. Box 260, DK-4000 Roskilde, Denmark}

\newcommand{\angleb}[1]{\left\langle #1 \right\rangle}
\newcommand{\half}{\frac{1}{2}}
\newcommand{\nod}{\noindent}
\newcommand{\bfa}[1]{\mathbf{#1}}


\begin{abstract}
We derive exact results for the rate of change of thermodynamic 
quantities, in particular the configurational
specific heat at constant volume, $C_V$,
 along configurational adiabats (curves of constant excess 
entropy $S_{\textrm{ex}}$). Such curves are designated isomorphs for so-called
Roskilde liquids, in view of the invariance of 
various structural and dynamical quantities along them. Their slope in a 
double logarithmic representation of the density-temperature phase diagram,
$\gamma$ can be interpreted as one third of an effective inverse power-law 
potential exponent.
We show that in liquids where $\gamma$ increases 
(decreases) with density, the contours of $C_V$ have smaller (larger) slope 
than configurational adiabats. We clarify also the connection between $\gamma$
and the pair potential. A fluctuation formula for the slope of the 
$C_V$-contours is derived. The theoretical
results are supported with data from computer
simulations of two systems, the Lennard-Jones fluid and the Girifalco fluid.
The sign of $d\gamma/d\rho$ is thus a third key parameter in 
characterizing Roskilde liquids, after $\gamma$ and the virial-potential
energy correlation coefficient $R$. To go beyond isomorph theory we compare
invariance of a dynamical quantity, the self-diffusion coefficient along 
adiabats and $C_V$-contours, finding it more invariant along adiabats.
\end{abstract}

\maketitle

\section{Introduction}

\begin{figure}

\epsfig{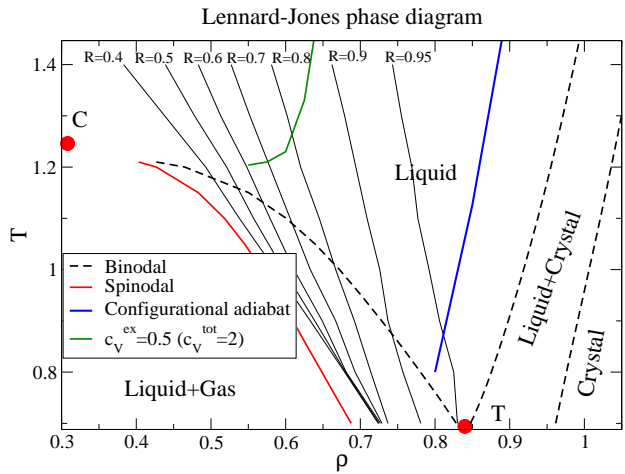}
\caption{\label{Rcontours}Contour plot of $R$ in $(\rho,T)$ phase diagram for the
single-component Lennard-Jones system using a shifted-potential cutoff 
of $4\sigma$ and system size $N=1000$. Contour values are indicated at the top. 
Also indicated are binodal and spinodal obtained from the Johnson equation of
state with the cutoff taken into account in a mean-field 
manner,\cite{Johnson/Zollweg/Gubbins:1993} and the corresponding curves for
solid-liquid coexistence as parameterized by Mastny and de Pablo (though for 
the larger cutoff 6$\sigma$).\cite{Mastny/dePablo:2007} T and C indicate the
triple\cite{Mastny/dePablo:2007} and 
critical\cite{Johnson/Zollweg/Gubbins:1993} points. The blue curve is a
configurational adiabat, while the green line is the configurational isochoric
specific heat contour $C_V=Nk_B/2$ (total specific heat $2Nk_B$); this is one 
of the criteria for the dynamic crossover separating liquid and gas regions
in the phase diagram proposed by Brazhkin et al.\cite{Brazhkin/others:2012a, Brazhkin/others:2012b, Brazhkin/others:2013} According to the theory of isomorphs 
both the configurational adiabat and the $C_V$-contour are
isomorphs for sufficiently large $R$.}
\end{figure}

The traditional notion of a simple liquid--involving point-like particles 
interacting via radially symmetric pair potentials\cite{fis64,ric65,tem68,ail80,rowlinson, gubbins,chandler,barrat,deb05,Hansen/McDonald:2005,dou07,kir07, bag10} (for example the Lennard-Jones (LJ )system)--is challenged by the existence 
of examples like the Gaussian core model,\cite{Stillinger:1976} and the 
Lennard-Jones Gaussian model\cite{Engel/Trebin:2007, VanHoang/Odagaki:2008} 
which exhibit complex behavior. Moreover many molecular models have simple 
behavior in computer simulations, and experiments on van der Waals liquids show 
that these are generally regular with no anomalous behavior.
We have recently 
suggested redefining a simple liquid--termed now a Roskilde-simple liquid, or
just a Roskilde liquid---as one with strong correlations 
between the equilibrium virial ($W$) and potential-energy ($U$)
fluctuations in the canonical fixed-volume
(NVT) ensemble.\cite{Ingebrigtsen/Schroder/Dyre:2012a} 
The basic phenomenology and theoretical understanding
of Roskilde liquids were presented in a series
of five papers published in the Journal of Chemical Physics. \cite{Bailey/others:2008b, Bailey/others:2008c,Schroder/others:2009,Gnan/others:2009,Schroder/others:2011} In particular, Appendix A of Ref.~\onlinecite{Gnan/others:2009} 
established an essential theorem of Roskilde liquids: A system has strong
$U,W$ correlations if and only if it has good isomorphs 
(curves in the thermodynamic phase diagram along which a number of properties 
are invariant in reduced units \cite{Gnan/others:2009}).
The degree of simplicity
depends on the thermodynamic state point---all realistic
systems lose simplicity when approaching the critical point and gas states. To 
illustrate this, Figure~\ref{Rcontours} shows the Lennard-Jones diagram 
including contours of the correlation coefficient $R$ between $U$ and $W$.
We choose an (arbitrary) cut-off $R>0.9$ as the boundary of simple-liquid 
behavior. It is clear from the figure that the correlation coefficient 
decreases rapidly as the liquid-gas spinodal is approached.

The theory of isomorphs starts with their definition and 
derives consequences from this which can be tested in simulations. For a 
system with $N$ particles, two density-temperature
state points $(\rho_1,T_1)$ and $(\rho_2, T_2)$ are 
isomorphic to each other if the Boltzmann factors for corresponding
configurational microstates are proportional:

\begin{equation}
\exp\left(-\frac{U(\bfa{r_1}^{(1)}, \ldots, \bfa{r_N}^{(1)})}{k_BT_1}\right)
 = C_{12} \exp\left(-\frac{U(\bfa{r_1}^{(2)}, \ldots, \bfa{r_N}^{(2)})}{k_BT_2}\right)
\end{equation}
Here $U$ is the potential energy function and $C_{12}$ depends on the two state
points, but not on which microstates are considered. Corresponding microstates 
means $\rho_1^{1/3} \bfa{r_i}^{(1)} = \rho_2^{1/3} \bfa{r_i}^{(2)}$, or 
$\bfa{\tilde r_i}^{(1)}=\bfa{\tilde r_i}^{(2)}$, where a tilde denotes so-called 
reduced units. Reduced units for lengths means multiplying by $\rho^{1/3}$, 
for energies dividing by $k_BT$, and for times dividing by 
$(m/k_BT)^{1/2}\rho^{-1/3}$ (for Newtonian dynamics). An isomorph is
a curve in the phase diagram of points which are isomorphic to each other.
From the definition it follows that all structural and dynamical correlation
functions are isomorph invariant when expressed in reduced units. Thermodynamic
quantities which do not involve volume derivatives, such as the excess entropy
$S_{\textrm{ex}}$ and excess specific heat at constant volume $C_V^{\textrm{ex}}$,
are also isomorph invariant. Another consequence of the isomorph definition is
that phase boundaries lying within the simple region of the phase diagram are
isomorphs---note that the isomorph shown in Fig.~\ref{Rcontours} is nearly
parallel to
the liquid-solid coexistence lines. Reference~\onlinecite{Pedersen/others:2011} 
gives a brief review of the theory and its experimentally relevant consequences.

Only inverse power-law (IPL) systems, i.e., systems for which
the potential energy is an Euler homogeneous function, have 100\% virial 
potential-energy correlation and perfect isomorphs. Thus for realistic Roskilde
liquids the isomorph concept is only approximate. Extensive computer simulations
have shown, however, that the predicted isomorph invariants apply
to a good approximation for several systems.
\cite{Ingebrigtsen/Schroder/Dyre:2012a,Gnan/others:2009, Veldhorst/others:2012,
  Ingebrigtsen/Schroder/Dyre:2012b,Schroder/others:2011,Gnan/others:2010,
  Ingebrigtsen/others:2012,Boehling/others:2013a} A few 
predictions have also been confirmed 
experimentally.\cite{Boehling/others:2013a, Gundermann/others:2011}

Despite the success of the isomorph concept, it remains a ``zero-order'' theory,
analogous to the ideal gas. In particular there is so systematic theory for
describing realistic systems in terms of perturbations about the ideal case. 
The purpose of this work is to examine deviations from perfect 
isomorph behavior in Roskilde liquids. One motivation is to 
understand what kind of deviations from 
IPL behavior (for example constancy of the scaling exponent) are allowed while 
remaining in the ``simple part'' of the phase diagram.
A second motivation is the hope of using Roskilde liquids to identify a general 
theory of liquids. For example, the existence of good isomorphs
explains
many observed connections between dynamics, structure and thermodynamics, but
also means that cause-and-effect interpretations of such connections (``the 
dynamics is {\em controlled} by \ldots'') must be re-examined. Given perfect
isomorphs, any isomorph-invariant quantity can be said to
control all the others. This puts a constraint on general theories, referred to
as the ``isomorph filter'',\cite{Gnan/others:2009} but prevents one from sorting
among theories that pass the filter. Examining carefully whether dynamical
properties are more invariant when holding one isomorph-invariant quantity 
fixed  versus holding another fixed could provide a means to select theories.

Strong $U,W$ correlation in the equilibrium 
NVT ensemble is a hallmark, and the first identified
feature,\cite{Pedersen/others:2008} of Roskilde liquids. It is characterized
at the level of second moments by the correlation coefficient 

\begin{equation}
R(\rho, T) =  \frac{\angleb{\Delta U\Delta W}}{\sqrt{\angleb{(\Delta U)^2}\angleb{(\Delta U)^2}}}
\end{equation}
and the slope 

\begin{equation}\label{gamma_definition}
\gamma(\rho, T) = \frac{\angleb{\Delta U\Delta W}}{\angleb{(\Delta U)^2}}.
\end{equation}
Here $\Delta$ represents the deviation of a quantity from its NVT ensemble
average. 
It has been shown that $\gamma$ may be thought of in terms of an effective 
inverse power-law (IPL) potential with exponent 3$\gamma$ (which in general
depends on state point).\cite{Bailey/others:2008b, Bailey/others:2008c} It has
also a thermodynamic interpretation, namely it is the ratio of the excess
pressure coefficient $\beta_V^{\textrm{ex}}\equiv (1/V)(\partial W/\partial T)_V$
and excess specific heat per unit volume,

\begin{equation}
\gamma = \frac{\beta_V^{\textrm{ex}}}{c_V^{\textrm{ex}}}.
\end{equation}
 
As mentioned, in IPL systems
the correlation is indeed perfect, but non-IPL systems exist which
yet have strong $U,W$-correlations, in 
particular the usual LJ fluid. While
in any system the fluctuation formula for $\gamma$ can be used to generate 
curves of constant (excess) 
entropy $S_{\textrm{ex}}$ (configurational adiabats) via\cite{Gnan/others:2009}

\begin{equation}\label{adiabats_from_gamma}
\left( \frac{\partial \ln T}{\partial \ln\rho} \right)_{S_\textrm{ex}} = \gamma (\rho, T),
\end{equation}
in Roskilde-simple liquids several properties related to structure,
thermodynamics, and dynamics are invariant along these curves. This leads to 
their designation as ``isomorphs''; note that quantities must be expressed in 
thermodynamically reduced units to exhibit the invariance. 
\cite{Gnan/others:2009} One of the most basic
isomorph-invariant quantities is the specific heat at constant volume: 
perfect isomorphs are also $C_V$-contours, while in imperfectly 
correlating systems the $C_V$ contours and configurational adiabats may differ. 

One might expect that the closer $R$ is to unity, the better approximated the
system would be by a single IPL potential. So it is perhaps surprising that we 
have recently identified systems
where $\gamma$ changes much more than in the LJ case, over a range
in which strong $U,W$-correlation ($R>0.9$) is maintained. 
One such system is the ``repulsive Lennard-Jones'' potential, in 
which the sign of the $1/r^6$ term is made 
positive.\cite{Ingebrigtsen/others:2012} It seems that the property of
strong $U,W$ correlation, and the existence of isomorphs are somehow more 
robust than the constancy of $\gamma$. It can be surprising how
well isomorphs ``work'' for non-IPL systems. This robustness allows for a 
richer variety
of behavior, since the shapes of isomorphs are no longer necessarily straight
lines in a $(\ln\rho,\ln T)$-plot. The theory of the thermodynamics of
Roskilde-simple liquids \cite{Ingebrigtsen/others:2012} implies that 
$\gamma$ may be considered a function of $\rho$ only. This immediately gives us
a new quantity (in addition to $R$ and $\gamma$) 
to characterize Roskilde liquids: $d\gamma/d\rho$, 
or more simply, its sign.  This result depends on
the assumption that configurational adiabats and $C_V$-contours exactly
coincide. It is not clear what to expect when this does not hold exactly; this
paper is an attempt to address the topic of imperfect correlation from
statistical mechanical considerations. Because $C_V$ is also a 
fundamental thermodynamic quantity, the difference between adiabats and 
$C_V$-contours should be a useful probe of the
breakdown of perfect isomorphs as $U,W$-correlation becomes less than perfect,
and will be the focus of this paper.

While as mentioned above, the arguments 
of Ref.~\onlinecite{Ingebrigtsen/others:2012} (which assume perfect isomorphs)
show that $\gamma=\gamma(\rho)$,
in practice $\gamma$ does depend on $T$ but the dependence is much
smaller than that on $\rho$, and we can ignore it most of the time. This 
is apparent for the single-component LJ system in Fig. 5 of
 Ref.~\onlinecite{Bailey/others:2008b}. A more explicit quantitative 
comparison, of the logarithmic derivatives of $\gamma$
with respect to $\rho$ and $T$, was made in 
Ref.~\onlinecite{Boehling/others:2013a} for two molecular systems. We present
further data on this below. Fluids with LJ and
similar potentials (for example generalized-LJ potentials with different 
exponents) tend to have $d\gamma/d\rho < 0$: It is clear that $\gamma$ must 
converge to one third of the repulsive exponent at very high densities
and temperatures while typical
values are larger.\cite{Bailey/others:2008c} 
 On the other hand potentials may be constructed
which have $d\gamma/d\rho > 0$, simply by shifting the potential radially 
outwards so that the repulsive 
divergence occurs at a finite value of pair separation.
Such potentials naturally involve a hard core of absolutely excluded volume.
They are relevant to experiments,\cite{Boehling/others:2013a} because
tests of the isomorph theory\cite{Gundermann/others:2011} typically 
involve molecules rather than single atoms, with the interaction range being 
relatively short compared to the particle size (colloids are of an even more 
extreme example of this, of course).
The Dzugutov system,\cite{Dzugutov:1992}
although only Roskilde-simple at high densities and temperatures,
is another example with $d\gamma/d\rho > 0$, but where there is no hard core.
Another such system is the above-mentioned repulsive Lennard-Jones
potential; in this case the effective exponent increases
monotonically, interpolating between the low density limit 6  ($\gamma=2$)
and the high density limit 12 ($\gamma=4$).

For brevity we term curves of constant $S_{\textrm{ex}}$  {\em adiabats} 
(the qualifier ``configurational'' is to be understood); in this paper, unlike
all our other works on isomorphs, we deliberately 
avoid calling them isomorphs, since the point of this work is to examine 
deviations from perfect isomorph behavior. We also drop the subscript 
$\textrm{ex}$ for notational simplicity, and similarly use $C_V$ to mean the 
configurational part of specific heat 
(the kinetic part is also isomorph invariant, though, being 3/2
for a classical monatomic system).
Below we derive some exact results concerning the relation between
adiabats and $C_V$-contours, and argue how this 
connects to whether $\gamma$ is an increasing or decreasing function of $\rho$
(more specifically the sign of $(\partial\gamma/\partial\rho)_S$). The argument
involves relating $\gamma$ to an exponent determined by derivatives of the 
pair potential, introduced in Ref.~\onlinecite{Bailey/others:2008c}. 
The claim is supported by simulations of two Roskilde liquids: the LJ fluid 
(with $d\gamma/d\rho<0$) and the
Girifalco fluid (with $d\gamma/d\rho>0$ at least for high densities). 
The Girifalco potential was constructed 
to model the C$_{60}$ molecules as spheres containing a uniform density
 of Lennard-Jones particles on their surface. Rotationally averaging gives the
following C$_{60}$-C$_{60}$ pair interaction\cite{Girifalco:1992}

\begin{equation}
v(r) = -\alpha\left(\frac{1}{s(s-1)^3} + \frac{1}{s(s+1)^3}-\frac{2}{s^4}\right)
+\beta\left(\frac{1}{s(s-1)^9} + \frac{1}{s(s+1)^9} -\frac{2}{s^{10}}\right).
\end{equation}

We have chosen the parameters $\alpha$ and $\beta$ such that the potential well
has a depth of approximately 1 and the potential diverges at unit distance,
$\beta = 0.0018141\alpha$ with $\alpha=0.17$.

For simulations we use systems of 1000 
particles simulated at constant volume and temperature (NVT) using
 the RUMD code\cite{RUMD:2012} for simulating on NVIDIA graphical 
processing units (GPUs). Although the state points considered do not involve
long relaxation times, the speed provided by GPUs is desirable because 
reasonably accurate determination of third moments requires of order one 
million independent samples; we typically run 50 million steps and sample every
50 steps (the time step sizes were 0.0025-0.004 for LJ and 0.0004 for 
Girifalco). The
temperature was controlled using a Nos\'e-Hoover thermostat.
Part (d) in Fig.~\ref{potentials_n2} shows the correlation coefficient $R$ along
an adiabat for each system. Both systems are Roskilde-simple (have
$R>0.9$) in the simulated part of the phase diagram.

In Section~\ref{thermo_derivs} a general fluctuation formula for derivatives of
 thermodynamic quantities along adiabats is derived, and applied to the case of
$C_V.$ In Section~\ref{sign_dCV_sign_dgamma} we show the connection between
the derivative of $C_V$ and derivatives of $\gamma$. The results are illustrated
with data from simulations. In Section~\ref{fluc_CV_contours} a fluctuation
formula for the slope of contours of $C_V$ is derived, and illustrated with
simulation data. The final two sections are the discussion and a brief 
conclusion.

\section{\label{thermo_derivs}Thermodynamic derivatives at constant entropy}

\subsection{$\gamma$ as linear-regression slope}

Before proceeding to thermodynamic derivatives we 
recall the connection between the above definition of $\gamma$ and linear
regression. Following Appendix C of Ref.~\onlinecite{Gnan/others:2009} we
characterize the deviation from perfect correlation via the fluctuating variable
\begin{equation}
\epsilon \equiv \Delta W-\gamma \Delta U,
\end{equation} 
which vanishes for perfect correlation. The linear regression slope is defined 
by minimizing $\angleb{\epsilon^2}$ with respect to $\gamma$, leading to 
Eq.~\eqref{gamma_definition}.\cite{Robbins/vanRyzin:1975} A consequence of this
 definition of $\gamma$ is seen by writing

\begin{equation} \label{W_U_epsilon}
\Delta W = \gamma \Delta U + \epsilon,
\end{equation}
and correlating\footnote{To ``correlate with'' means to multiply by and take an 
ensemble average.} this with $\Delta U$:

\begin{equation}
\angleb{\Delta W \Delta U} = \gamma \angleb{(\Delta U)^2}  + \angleb{\Delta
 U\epsilon}
\end{equation}
From this and the definition of $\gamma$ it follows that

\begin{equation}\label{corr_DeltaU_epsilon}
\angleb{\Delta U\epsilon}=0,
\end{equation}
that is, $U$ and $\epsilon$ are (linearly) uncorrelated,
independent of whether perfect correlation holds between $U$ and $W$. 

\subsection{Density-derivatives of averages on adiabats}

We are interested in the derivatives of thermodynamic quantities along certain
curves in the phase diagram, in particular those of constant $S$, 
so we start by presenting general formulas for
the derivatives with respect to $\ln\rho$ and $\ln T$ 
(holding the other constant). From standard statistical mechanics (see, for
example, appendix B of Ref.~\onlinecite{Bailey/others:2008b}) we have
(with $\beta=1/(k_BT)$; in the following we set $k_B=1$)

\begin{equation}\label{dA_d_beta}
\left( \frac{\partial \angleb{A}}{\partial\beta}\right)_\rho = -\angleb{\Delta U\Delta A}
\end{equation}
which implies

\begin{equation}\label{dA_d_ln_T}
\left( \frac{\partial \angleb{A}}{\partial\ln T}\right)_\rho = \beta\angleb{\Delta U\Delta A}.
\end{equation}
Likewise (see appendix \ref{dA_d_ln_rho_deriv} )

\begin{equation}\label{dA_d_ln_rho}
\left(\frac{\partial\angleb{A}}{\partial \ln\rho}\right)_T
= \angleb{\frac{\partial A}{\partial \ln\rho}}
-\beta\angleb{\Delta W\Delta A},
\end{equation}
where differentiation with respect to $\ln\rho$ inside an expectation 
value---that
is, for an arbitrary configuration rather than an ensemble average---is
understood to imply that the reduced coordinates of the configuration,
$\tilde{\bfa{r}}_i\equiv \rho^{1/3}\bfa{r}_i$, are held fixed. 
Equations~\eqref{dA_d_ln_T} and \eqref{dA_d_ln_rho} can be used to construct 
the derivative with respect to $\ln\rho$ along an arbitrary direction; that is 
instead of keeping $T$ constant (a line of zero slope) we take a direction with
 slope $g$ (in $\ln\rho, \ln T$ space):

\begin{align}
\left( \frac{\partial \angleb{A}}{\partial\ln\rho}\right)_{[g]} &= 
\left(\frac{\partial \angleb{A}}{\partial\ln\rho}\right)_T + 
g \left(\frac{\partial \angleb{A}}{\partial\ln T}\right)_\rho \\
&= \angleb{\frac{\partial A}{\partial \ln\rho}}
-\beta\angleb{\Delta W\Delta A} + g \beta\angleb{\Delta U\Delta A}\\
&=  \angleb{\frac{\partial A}{\partial \ln\rho}}
-\beta\angleb{\Delta A(\Delta W -g\Delta U)}
\end{align}
Note that we use subscript $[g]$ to indicate that $g$ is the slope in the 
$\ln\rho, \ln T$ plane, rather than the quantity held constant, in the 
derivative.
This expression can be used to find formulas for the direction in which a given
thermodynamic variable is constant, as we do below. For now we choose
$g=\gamma$, to obtain a formula for derivatives along adiabats 
(Eq.~\eqref{adiabats_from_gamma}):

\begin{equation}\label{deriv_A_const_S}
\left(\frac{\partial \angleb{A}}{\partial \ln\rho} \right)_S =
 \angleb{\frac{\partial A}{\partial\ln\rho}} 
- \beta \angleb{\Delta A \Delta(W-\gamma U)}  = 
 \angleb{\frac{\partial A}{\partial\ln\rho}} - \beta\angleb{\Delta A\ \epsilon}
\end{equation}

As an example, we take $A=U$. Noting that
$W\equiv \partial U/\partial\ln\rho$ and
Eq.~\eqref{corr_DeltaU_epsilon}, we get

\begin{equation}\label{dU_d_rho_S}
\left(\frac{\partial \angleb{U}}{\partial\ln\rho}\right)_S = 
\angleb{\frac{\partial U}{\partial\ln\rho}}=\angleb{W},
\end{equation}
which is a general result that can also be derived thermodynamically starting
with the fundamental thermodynamic identity $T dS=dU+pdV=dU + W d\ln(V) = 
dU-Wd\ln\rho$ (here the variables $U$, $W$ refer to macroscopic, or thermally
averaged quantities, the omission of angle-brackets notwithstanding). 
As a second application of Eq.~\eqref{deriv_A_const_S}, consider
a system with perfect correlation. Then $\epsilon \equiv 0$, and
we get 

\begin{equation}\label{deriv_A_const_S_R_equals_one}
\left(\frac{\partial \angleb{A}}{\partial \ln\rho} \right)_S =
 \angleb{\frac{\partial A}{\partial\ln\rho}},
\end{equation}
which means that in such systems the derivative along an adiabat
 is given entirely by the ``intrinsic'' density dependence for individual
configurations; fluctuations do not contribute. This is of course the case of
perfect isomorphs, where the probabilities of scaled configurations
are identical along an isomorph.

\subsection{\label{VariationCV}Variation of $C_V$ on adiabats}

We consider the derivative of $C_V$ with respect to $\ln\rho$ on an adiabat.
From $C_V=\angleb{(\Delta U)^2}/T^2$, we have

\begin{align}
\left( \frac{\partial C_V}{\partial \ln\rho}\right)_S &= 
\frac{1}{T^2} \left(\frac{\partial \angleb{(\Delta U)^2} }
{\partial\ln\rho}\right)_S
 - \frac{2}{T^3} \angleb{(\Delta U)^2}
 \left(\frac{\partial  T}{\partial \ln\rho}\right)_S \label{d_Cv_d_rho_S_1} \\
&=   \frac{1}{T^2} \left(\frac{\partial\angleb{(\Delta U)^2} }{\partial
\ln\rho}\right)_S   - \frac{2\gamma}{T^2} \angleb{(\Delta U)^2}
 \label{d_Cv_d_rho_S_2}
\end{align}

Writing $\angleb{(\Delta U)^2}= \angleb{U^2}-\angleb{U}^2$ and
making use of the general result of Eq.~\eqref{deriv_A_const_S}, after some
algebra (see appendix~\ref{Derivation_C_V_deriv}) we obtain the simple result

\begin{equation}\label{Deriv_CV_curve_const_S}
\left( \frac{\partial C_V}{\partial \ln\rho}\right)_S =
 - \beta^3 \angleb{(\Delta U)^2\Delta(W-\gamma U)} = -\beta^3\angleb{(\Delta U)^2\epsilon}
\end{equation}
This is a major result of this paper. Note that the right side vanishes
for perfect correlation ($\epsilon=0$)---in which case $C_V$ is constant on the 
same curves that $S$ is; in other words,
 $C_V$ is a function of entropy only. For less than perfect correlation, the
most interesting feature is the sign, which we argue in the next section, is the
opposite of that of $d\gamma/d\rho$.

\section{\label{sign_dCV_sign_dgamma}Connection between  $(\partial C_V/\partial\rho)_S$ and derivatives of $\gamma$}

\subsection{Relation to temperature-dependence of $\gamma$}

We wish to understand the sign of $\angleb{(\Delta U)^2\epsilon}$.
We know from Eq.~\eqref{corr_DeltaU_epsilon} that $U$ and $\epsilon$ are
linearly uncorrelated; we must now consider higher order correlations.
Recall that $\gamma$ may also be interpreted\cite{Bailey/others:2008b}
as the slope of isochores in the $W,U$ phase diagram--the linear regression of
the scatter-plot of instantaneous $W,U$ values at one state point gives the 
slope of $\angleb{W}$ versus $\angleb{U}$ at fixed density. The triple
correlation is related to the curvature of the isochore, and thus
to $(\partial\gamma/\partial T)_\rho $. We can get the exact relation
differentiating $\gamma$ with respect to $\beta$:

\begin{align}
\left(\frac{\partial \gamma}{\partial \beta}\right)_\rho &= 
\frac{1}{\angleb{(\Delta U)^2}}
\left(\frac{\partial \angleb{\Delta U\Delta W}}{\partial \beta}\right)_\rho
- \frac{\angleb{\Delta U\Delta W}}{\angleb{(\Delta U)^2}^2} 
\left(\frac{\partial \angleb{(\Delta U)^2}}{\partial\beta}\right)_\rho \\
&= -\frac{\angleb{(\Delta U)^2\Delta W}}{\angleb{(\Delta U)^2}}
-\frac{\gamma}{\angleb{(\Delta U)^2}} \left(-\angleb{(\Delta U)^3}\right) \\
&= -\frac{\angleb{(\Delta U)^2(\Delta W-\gamma\Delta U)}}{\angleb{(\Delta U)^2}} \\
&= -\frac{\angleb{(\Delta U)^2\epsilon}}{\angleb{(\Delta U)^2}},
\end{align}
where we have used Eq.\eqref{dA_d_beta} and some algebraic manipulation as in
Appendix~\ref{Derivation_C_V_deriv}. Combining this result with 
Eq.~\eqref{Deriv_CV_curve_const_S} gives

\begin{equation}\label{d_C_V_d_rho_S_d_gamma_dT}
\left( \frac{\partial C_V}{\partial \ln\rho}\right)_S =
\beta^2\angleb{(\Delta U)^2}
\beta\left(\frac{\partial\gamma}{\partial\beta}\right)_\rho 
= -C_V \left(\frac{\partial\gamma}{\partial \ln T}\right)_\rho,
\end{equation}
or more concisely
\begin{equation}\label{d_C_V_d_rho_S_d_gamma_dT_concise}
\left( \frac{\partial \ln C_V}{\partial \ln\rho}\right)_S = 
-\left(\frac{\partial\gamma}{\partial \ln T}\right)_\rho.
\end{equation}

\subsection{Relation to density-dependence of $\gamma$ via the
effective IPL exponent $n^{(2)}(r)$}

The last result implies, in particular, that
 the sign of the density-derivative of $C_V$ along an isomorph is opposite to
that of  $(\partial\gamma/\partial T)_\rho$.
Since the latter derivative is neglected in 
the theory of isomorphs, it is useful to find a connection with a
density derivative of $\gamma$. The relevant derivative turns out not to be
$(\partial\gamma/\partial\rho)_T$ but $(\partial\gamma/\partial\rho)_S$, i.e. the
derivative of $\gamma$ along the 
adiabat. For many systems of interest this derivative
has the same sign as $(\partial\gamma/\partial T)_\rho$, while those signs 
can be positive or negative depending on the system (or even for a given 
system). We shall now argue that this sign-equivalence is to be expected by 
considering how $\gamma$ is related to the 
pair potential $v(r)$. This is an interesting question in its own right, and was
explored in Ref.\onlinecite{Bailey/others:2008c}. For potentials with strong 
repulsion at short distances, 
we can indeed relate $\gamma$ directly, albeit approximately, to $v(r)$, or 
more precisely, to its 
derivatives. As discussed in Ref.~\onlinecite{Bailey/others:2008c} the idea is
to match an IPL to the actual potential; $\gamma$ is then one third of the
``effective IPL exponent''. There are many ways to define such an exponent, but
a key insight is that
it should involve neither the potential itself (because 
shifting the zero of potential has no consequences), nor its first
derivative (because the contributions to the forces from a linear term tend to 
cancel out in dense systems at fixed volume).\cite{Bailey/others:2008c} The 
simplest possibility within these constraints involves the
ratio of the second and third derivatives. For an IPL,
 $v(r)\propto 1/r^n$, and indicating
derivatives with primes, we have 
$v'''(r)/v''(r) = -(n+2)/r$, so $n$ can be extracted as $-r v'''(r)/v''(r)-2$.
For a general pair potential this quantity will be a function of $r$, and thus 
we define the $r$-dependent second-order effective IPL exponent 
$n^{(2)}(r)$ as\cite{Bailey/others:2008c}

\begin{equation}
n^{(2)}(r) \equiv -\frac{r v'''(r)}{v''(r)} - 2
\end{equation}
The superscript ``(2)'' indicates which derivative appears in the denominator;
one can similarly\cite{Bailey/others:2008c} define $n^{(p)}(r)$ for $p=0,1, 
\ldots$; $p=2$ is the first not involving $v$ or $v'$. Interestingly, the
IPL is not the only solution to $n^{(2)}(r)=n$ with constant $n$; so is the 
so-called extended IPL

\begin{equation}\label{extendedIPL}
v_{\textrm{eIPL}}(r)=A/r^n + Br + C,
\end{equation}
introduced in Ref.\onlinecite{Bailey/others:2008c}. The resemblance of the
Lennard-Jones potential to such a form can be considered an explanation of why
it behaves like an IPL system. For a general potential, the question 
that now arises is at which $r$ one should evaluate $n^{(2)}$. It was
argued in Ref.~\onlinecite{Bailey/others:2008c} that $n^{(2)}/3$ evaluated at a 
point near the maximum of $g(r)$---let us call it $r_\gamma$---should correspond 
to $\gamma$. One expects that, like the peak in $g(r)$, 
$r_\gamma = \Lambda\rho^{-1/3}$, where $\Lambda$ is of order unity and depends
weakly on temperature, but we do not know it precisely {\em a priori}. 
There are two crucial things we can say, however: First,
we can certainly identify $r_\gamma$ {\em a posteriori} by 
inspection for a given state point: That is, having simulated a reference state
point $(\rho_{\textrm{ref}}, T_{\textrm{ref}})$ and determined $\gamma_\textrm{ref}$ 
there, it is straightforward to (typically numerically) solve the equation 
$n^{(2)}(r_{\gamma})/3=\gamma_{\textrm{ref}}$ for $r_\gamma$. The second crucial point is
 that whatever 
details of the liquid's statistical mechanics determine $r_\gamma$ (for instance
a kind of $g(r)$-weighted average), {\em these details do not vary along an
isomorph} (this argument assumes good isomorphs, so that the statement can be
applied to adiabats). Therefore $r_\gamma$ is an isomorph invariant---more 
precisely its reduced-unit form $\rho^{1/3} r_\gamma =\Lambda$ is constant along 
an adiabat, which implies $\Lambda = \Lambda(S)$. So
$\gamma$ is given by

\begin{equation}\label{gamma_rho_estimate}
\gamma(\rho, S) = 
\frac{1}{3} n^{(2)}( \Lambda(S) \rho^{-1/3} ),
\end{equation}
or
\begin{equation}\label{gamma_rho_estimate2}
\gamma(\rho, S) = 
\frac{1}{3} n^{(2)}\left( r_{\gamma,\textrm{ref}} (S)
\frac{\rho^{-1/3}}{\rho_{\textrm{ref}}^{-1/3}}\right).
\end{equation}
In the form with $\Lambda$ we explicitly recognize that $\Lambda$ is 
constant on an isomorph, or equivalently, that it depends on $S$; the second
form shows how $\Lambda$ can be determined using a simulation at one
density to identify $r_\gamma$ there.

For the 
Lennard-Jones potential $n^{(2)}(r)$ decreases as $r$ decreases
(corresponding to as $\rho$ increases), while
for potentials such as the Girifalco potential with a divergence at finite $r$
(see Fig.~\ref{potentials_n2} below), it increases as $r$ decreases ($\rho$ 
increases), although at low densities the opposite behavior is 
seen. The validity of Eq.~(\ref{gamma_rho_estimate}) has been investigated by
B{\o}hling et al.\cite{Boehling/others:2013b} Under which circumstances does
Eq.~\eqref{gamma_rho_estimate} give a good estimate of the density
dependence of $\gamma$? The system must have sufficiently strong $W,U$ 
correlations, since as $R\rightarrow0$, $\gamma$ must also vanish irrespective 
of $n^{(2)}$'s 
behavior. (For example, in a Lennard-Jones-like 
liquid, as $r$ increases, the curvature of the pair potential 
becomes negative at some $r$, at which point $n^{(2)}$ diverges. At or below
the corresponding density, and not too high temperature, a single phase is 
likely to have a negative pressure and be mechanically unstable, giving way to 
liquid-gas coexistence. In this regime $W,U$ correlations tend to break 
down completely and $\gamma$ goes to zero; see Fig.~\ref{potentials_n2} (c) and
 (d) in particular the Girifalco data.)

\begin{figure}
\epsfig{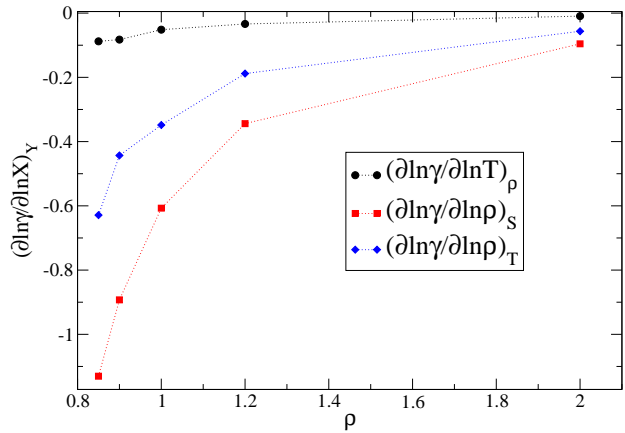}
\caption{\label{CompareGammaDerivsLJ} Logarithmic derivatives of $\gamma$:
(1) with respect to $T$ at constant $\rho$,
(2) respect to $\rho$ at constant $S$ and (3) respect to $\rho$ at constant
$T$, for the LJ system at points along the adiabat including $\rho=0.85$, 
$T=0.80$. The first derivative
was determined via fitting $\ln(\gamma)$ versus $\ln T$ data (obtained 
also for neighboring adiabats) at each $\rho$ to a quadratic
function; the second analytically after making a (one-parameter) fit to the 
logarithmic derivative of  Eq.~\eqref{h_of_rho}, and the third via the chain 
rule as a linear combination of the other two, 
$(\partial\ln(\gamma)/\partial\ln\rho)_T = 
(\partial\ln(\gamma)/\partial\ln\rho)_S
-\gamma (\partial\ln(\gamma)/\partial\ln T)_\rho$.
While all decrease to zero at high densities (consistent
with $\gamma$ converging to a constant 4=12/3) the temperature derivative is 
consistently a factor of ten smaller than the density derivative at constant
$S$. }
\end{figure}

Equation~\eqref{gamma_rho_estimate} shows how $\gamma$ depends on $\rho$, but
we need to consider temperature dependence in order to connect with the result
for $C_V$ along an adiabat. This comes in through $\Lambda(S)$. We cannot right
away determine how $\Lambda$ depends on $S$ but we know it is a weak dependence,
since $r_\gamma$ is expected to remain close to the peak in 
$g(r)$.\cite{Boehling/others:2013b} For liquids with a repulsive core this 
peak moves slowly to shorter distances as temperature, and hence entropy, 
increase at fixed $\rho$. We expect the same to be true for $\Lambda$, since
in the high-temperature limit potential energy and virial fluctuations, and 
thus $\gamma$, are dominated by ever 
smaller pair separations. Thus we expect that

\begin{equation}\label{d_Lambda_d_S_negative}
\frac{d\Lambda (S)}{dS} < 0,
\end{equation}
while the weak dependence on entropy/temperature at fixed density can be 
expressed as

\begin{equation}\label{d_LogLambda_d_S_small}
C_V \frac{d\ln\Lambda (S)}{dS} \ll 1,
\end{equation}
(the use of $C_V$ to make the left side dimensionless, instead of for 
example. differentiating with respect to $\ln S$, is done for convenience below;
note that $C_V$ varies slowly and has 
a similar order of magnitude to the entropy differences between isomorphs in the
liquid region of the phase diagram).
From Eq.~\eqref{d_Lambda_d_S_negative} it follows that both increasing $\rho$
at fixed $S$, and increasing $T$ at fixed $\rho$, decrease the argument of 
$n^{(2)}$. (Recall that in the earliest work on
Roskilde liquids it was noted that the slope of the $W,U$ 
correlation converges down towards 12/3=4 for the LJ case both in the high 
temperature and the high density limits.\cite{Pedersen/others:2008})
Taking the appropriate derivatives of Eq.~\eqref{gamma_rho_estimate} yields

\begin{align}
\left(\frac{\partial\gamma}{\partial\ln\rho}\right)_S &= 
-\frac{\Lambda(S) \rho^{-1/3}}{9} 
\left.\frac{d n^{(2)}(r)}{d r}\right|_{r=\Lambda(S)\rho^{-1/3}} \\
\left(\frac{\partial\gamma}{\partial\ln T}\right)_\rho &= 
\frac{\Lambda(S) \rho^{-1/3}}{3} 
\left.\frac{d n^{(2)}(r)}{d r}\right|_{r=\Lambda(S)\rho^{-1/3}} \frac{d\ln(\Lambda(S))}{d S} C_V.
\end{align}
Combining these gives
\begin{equation}
\left(\frac{\partial\gamma}{\partial\ln T}\right)_\rho =
\left(\frac{\partial\gamma}{\partial\ln\rho}\right)_S  
\left( -3 \frac{d\ln(\Lambda(S))}{d S} C_V   \right), 
\end{equation}
From Eqs.~\eqref{d_Lambda_d_S_negative} and \eqref{d_LogLambda_d_S_small} 
the quantity in brackets on the right side is positive but much smaller than 
unity. We therefore have

\begin{equation}\label{gamma_derivatives_conjecture}
\sgn\left( \left(\frac{\partial\gamma}{\partial T}\right)_\rho \right)
= \sgn\left( \left(\frac{\partial\gamma}{\partial\rho}\right)_S \right) , \quad
\left| \left(\frac{\partial\gamma}{\partial T}\right)_\rho \right| \ll 
\left|\left(\frac{\partial\gamma}{\partial\rho}\right)_S \right|,
\end{equation}
which is expected to hold for liquids with repulsive cores, with sufficiently
strong $W,U$-correlations. It remains to be investigated thoroughly to what
extent Eq.~\eqref{gamma_derivatives_conjecture} holds, both 
regarding in how large a region of the phase diagram it holds for a given 
liquid, 
and for which liquids it holds in a reasonably large region. Its validity 
depends both on that of Eq.~\eqref{gamma_rho_estimate} and the conjecture that
$\Lambda$ decreases, slowly, as entropy increases. Some data 
is shown in Table~\ref{conjecture_data}
which compares the signs of the two derivatives for different systems and
Fig.~\ref{CompareGammaDerivsLJ} which compares the two derivatives at state
points along an adiabat for the LJ system. For comparison the density 
derivative at fixed temperature is also shown, obtained via chain-rule 
combination
of the other two derivatives. This involves a minus sign and therefore the two 
terms (which have the same sign) tend to cancel.

\begin{table}
\begin{tabular}{|c|c|c|c|c|}
\hline 
Potential & $\rho$-range & $T$-range & $(\partial\gamma/\partial T)_\rho$ &
$(\partial\gamma/\partial \rho)_S$  \\
\hline
Lennard-Jones & 0.6-1.2 & 0.8-5.0 & - & - \\
Buckingham & 0.7-1.2 & 2-6 & - & -\\
Dzugutov & 0.55-0.8 & 0.75-1.2 & + & + \\
Girifalco & 0.45-0.5 & 6-54 & + & + \\
Repulsive Lennard-Jones& 0.1-10 &0.4-2.0  & + & + \\
\hline
\end{tabular}
\caption{\label{conjecture_data} Validity of 
Eq.~\eqref{gamma_derivatives_conjecture} for several potentials. For each system
the signs of $(\partial\gamma/\partial T)_\rho$ and 
$(\partial\gamma/\partial \rho)_S$ have been checked for a set of adiabats. For
the Lennard-Jones, Buckingham and Dzugutov system the
density range gives the lowest densities of the simulated adiabats
while the temperature range gives the range of temperatures simulated for
each adiabat. For the Girifalco and 
repulsive Lennard-Jones the density range indicates the range of densities 
simulated for each adiabat, while the temperature range indicates the 
lowest temperatures. Data near extrema of $\gamma$ have not been included.
}
\end{table}

In the limit of perfect $W,U$ correlation we know 
$(\partial\gamma/\partial T)_\rho$ vanishes. There is no reason to expect 
$\Lambda$ to become constant in this limit,\footnote{Our method of determining
$\Lambda$ fails when $n^{(2)}$ is constant, but the value of $r$ where one 
should evaluate $n^{(2)}$ is in principle well-defined.} 
therefore $(\partial\gamma/\partial\rho)_S$ must also vanish in the limit. This 
corresponds to $n^{(2)}(r)$ becoming
constant: IPL or extended IPL systems (Eq.~\ref{extendedIPL}).
But because the dependence of $\Lambda$ on $S$ is
in general weak, there is a regime---that of general Roskilde liquids---where
we can neglect it, but where $n^{(2)}$ cannot be considered constant.
In this approximation, then, we can write the density derivative as an ordinary
derivative. Combining this with Eq.~\eqref{d_C_V_d_rho_S_d_gamma_dT_concise}
we have the following result for the sign of the $C_V$:

\begin{equation}
\sgn\left( \left(\frac{\partial C_V}{\partial \ln\rho}\right)_S\right) 
= - \sgn\left(d\gamma/d \rho\right).
\end{equation}

Thus we can predict---based 
on the $n^{(2)}$ estimate of $\gamma$---that the rate of change of $C_V$ along
an adiabat has the opposite sign as the density dependence of $\gamma$ 
(along the adiabat if we need to be specific). 
Thus from knowing only the pair potential one can say something reasonably
accurate about both the adiabats and the $C_V$-contours.

\subsection{Simulation results for variation of $C_V$ along adiabats}

\begin{figure}
\epsfig{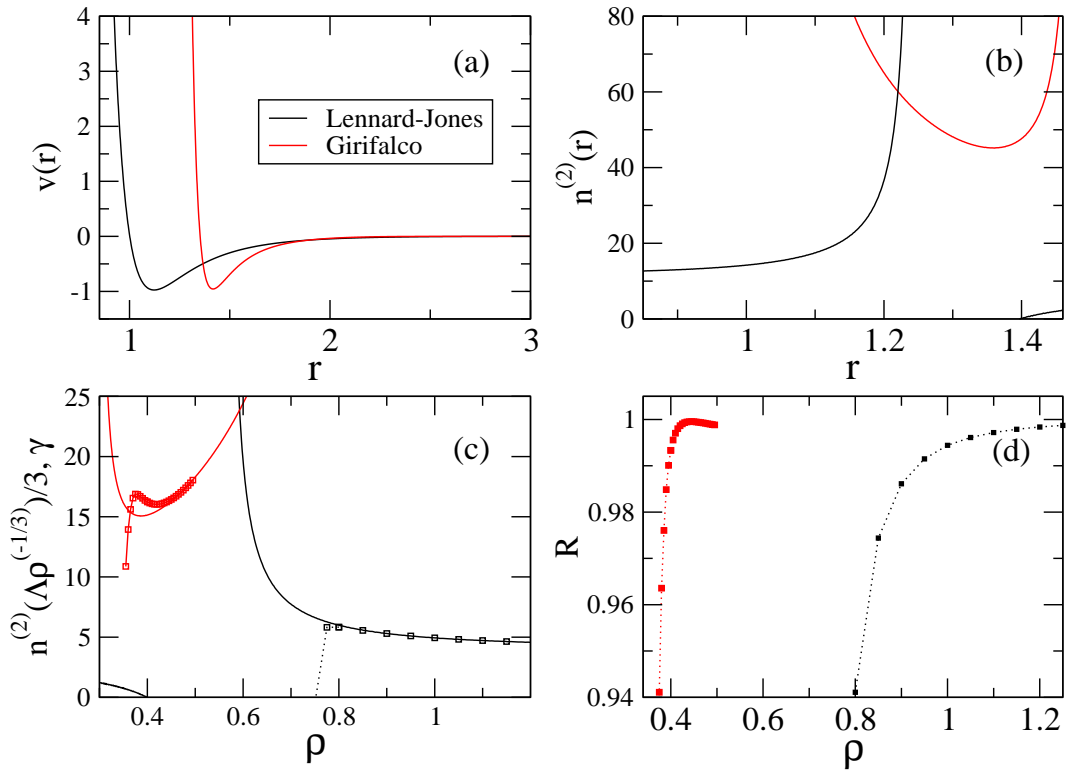}
\caption{\label{potentials_n2}
(a) The pair potentials used in this work. The Girifalco potential
diverges at $r=1$. (b) $n^{(2)}(r)$ for the two potentials.
(c) $n^{(2)}(\Lambda \rho^{-1/3})$ (full lines) and $\gamma$ on sample adiabats 
for both models (symbols). The entropy was not calculated, but
adiabats are uniquely specified by giving one state point, for example 
$\rho=0.80$, $T=0.80$ for the LJ case and $\rho=0.4$, $T=4.0$ for the GF case.
The value of $\Lambda$ was fixed by requiring agreement with $\gamma$ at the 
highest simulated density for each isomorph.
(d) Correlation coefficient $R$ from simulations,
along the same adiabats as in (c).}
\end{figure}

To confirm the relation between the sign of $d\gamma/d\rho$ and that of 
$(\partial C_V/\partial \rho)_S$ and exhibit the relation between 
adiabats and $C_V$ contours we carried out simulations on two model systems. 
Figure~\ref{potentials_n2}(a) shows the pair potentials.
Note that the Girifalco potential diverges at $r=1$; this hard core 
restricts the density to be somewhat smaller than for the LJ case, if a 
non-viscous liquid is to be considered.
Part (b) shows the effective exponent $n^{(2)}(r)$. There is a singularity where 
the second derivative vanishes (the transition from concave up to concave 
down), which can be seen in the figure at $r \simeq 1.224$ for LJ and 
$r\simeq 1.48$ for 
Girifalco; as 
 $r$ decreases from the singularity $n^{(2)}$ decreases monotonically in the
LJ case, while in the Girifalco case it first decreases and then has a minimum
 before increasing and in fact diverging as $r=1$ is approached. 
Part (c) of the figure shows the estimate of $\gamma(\rho)$ from 
Eq.~(\ref{gamma_rho_estimate}) along with $\gamma(\rho)$
calculated in simulations along an adiabat for each system. Here $\Lambda$ was
determined by matching $n^{(2)}/2$ with $\gamma$ at the highest density. 
The agreement is good for not too low densities---as mentioned above when 
$n^{(2)}(r)$ diverges due to the curvature of the potential vanishing, then
both $R$ and $\gamma$ will rapidly approach zero, which is what we can see
happening for the Girifalco system in parts (c) and (d) and low density.
Note that the adiabat for the Girifalco system rapidly reaches
rather high temperatures, since the exponent is always greater than 15, or 
roughly three times that of the LJ system. More interestingly, for the Girifalco
system $d\gamma/d\rho$ changes sign at a density around 0.4, so we can expect
the dependence of $C_V$ along an adiabat to reflect this. The location and 
value of the minimum in $\gamma$ do not match those for $n^{(2)}$, 
however---perhaps the vanishing of the curvature is already having an effect.



\begin{figure}
\epsfig{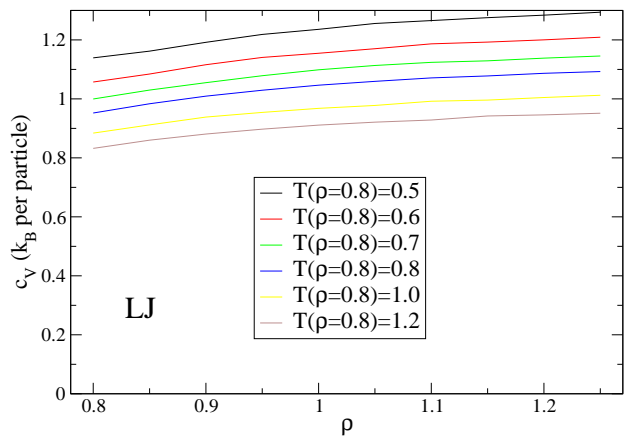}
\caption{\label{CV_along_iso_LJ} Dependence of $c_V=C_V/N$ on density along 
three different adiabats for the LJ fluid. We label the curves by their 
temperature 
at a fixed density, here the starting density $\rho=0.8$. The change
in $c_V$ is of order 0.1-0.15 for the $\sim$50\% change in density shown here,
small but not negligible. The slopes are positive, consistent with the negative 
sign of $d\gamma/d\rho$ and arguments of
Section~\ref{sign_dCV_sign_dgamma}.}
\end{figure}

\begin{figure}
\epsfig{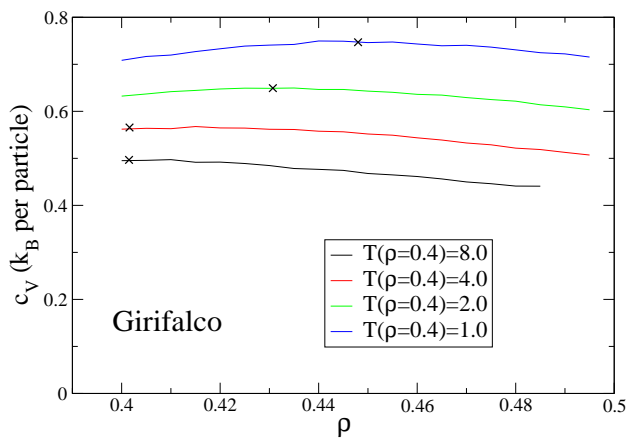}
\caption{\label{CV_along_iso_GF} Dependence of $c_V=C_V/N$ on density along four
 different adiabats for the Girifalco fluid. The curves
are labelled by their temperature at $\rho=0.4$. For the $\sim$20\% changes
in density shown here, $c_V$ changes by about 0.05. It is generally decreasing
in the range shown
but increases at low densities and temperatures; the maxima (determined by 
fitting a cubic polynomial) are shown as crosses, and appear at different 
densities for different adiabats.}
\end{figure}

The procedure for determining adiabats is described in 
Appendix~\ref{generating_adiabats}.
Figures~\ref{CV_along_iso_LJ} and \ref{CV_along_iso_GF} show $c_V=C_V/N$ along 
adiabats for the LJ and Girifalco systems, respectively.  For the LJ
case the slope is positive, which is consistent with $d\gamma/d\rho$ being
negative as discussed in Section~\ref{sign_dCV_sign_dgamma}. 
It is worth noting that
the overall variation of $C_V$ is quite small, of order 0.1 per particle
for the density range shown, but it is not negligible, even though the system
has strongly $U,W$ correlations and the structure and dynamics have been shown 
to be quite invariant along the adiabats. For the Girifalco system the slope is 
positive at low density until a maximum is reached, with a negative slope at
higher densities. This is also broadly consistent with the expectations from
Fig.~\ref{potentials_n2} (the locations of the maxima are not expected 
to be accurately given by Eq.~\eqref{gamma_rho_estimate}).



\subsection{Contours of $C_V$ and $S$ directly compared}

\begin{figure}
\epsfig{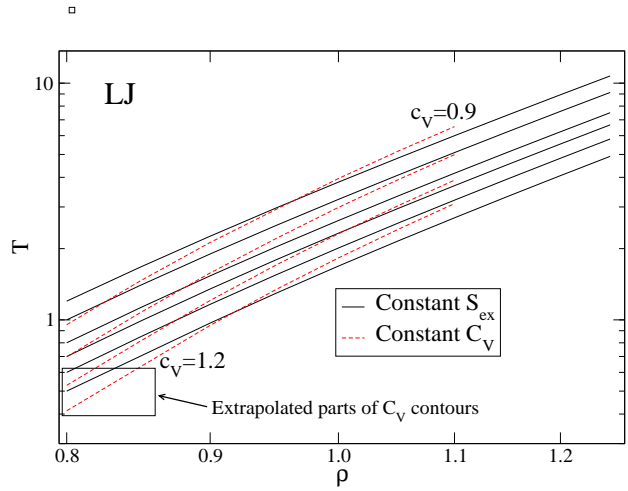}
\caption{\label{const_S_vs_const_CV_LJ}  Comparison of adiabats and
$C_V$-contours for the LJ system. The adiabats are the same as those shown in
Fig.~\ref{CV_along_iso_LJ}, and were calculated using 
Eq.~(\ref{long_jump_formula}), while $C_V$-contours (values 0.9, 1.0, 1.1
and 1.2 in units of $k_B$)
were determined from a series of simulations on different isochores and
interpolating the $C_V$ data as a function of $T$ (some extrapolated points, 
indicated, were also included). }
\end{figure}

\begin{figure}
\epsfig{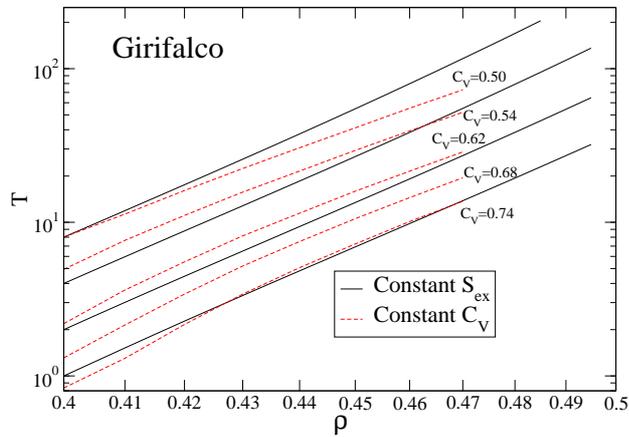}
\caption{\label{const_S_vs_const_CV_GF}  Comparison of adiabats with
$C_V$-contours for the Girifalco system. 
The adiabats were
calculated using the the definition of $\gamma$ and small changes in $\rho$, 
while $C_V$-contours 
were determined from a series of simulations on different isochores and
interpolating the $C_V$ data as a function of $T$.}
\end{figure}

As an alternative to considering how $C_V$ varies along an
adiabat, we can find the contours of $C_V$ separately. First we
simulated several isochores, then the data were interpolated to allow 
constant-$C_V$ curves to be constructed. Specifically, we find that the 
dependence of $C_V$ on temperature along an isochore can be accurately fitted
by the expression

\begin{equation}\label{CV_T_dep_fit}
C_V(T) = \frac{A(\rho)}{T^{B(\rho)}} + C(\rho),
\end{equation}
where $A$, $B$ and $C$ are functions of $\rho$. This expression was 
inspired by the Rosenfeld-Tarazona expression $C_V \sim T^{-2/5}$ for the
specific heat;\cite{Rosenfeld/Tarazon:1998} we do not constraint the exponent
$B$ to be 2/5, however. The expression can easily be inverted to yield the 
temperature $T_{C_V}(\rho)$ 
corresponding to a given value of $C_V$, as a function of density 

\begin{equation}
T_{C_V}( \rho ) = \left(\frac{A(\rho)}{C_V-C(\rho)} \right)^{1/B(\rho)},
\end{equation}
The $C_V$ contours are shown along with the adiabats in Figs.~\ref{const_S_vs_const_CV_LJ} and \ref{const_S_vs_const_CV_GF}. Recall that in typical
liquids we expect
$C_V$ to increase as $T$ decreases or $\rho$ increases. For the LJ case the
$C_V$ contours have a higher slope than the adiabats, therefore as $\rho$
increases along an adiabat we cross contours corresponding to higher 
values of $C_V$. For the Girifalco system the $C_V$ contours have initially 
(at low density) higher slopes than the adiabats but then bend over and have
lower slopes. Thus the picture is consistent with the data for $C_V$ along
adiabats shown in Figs.~\ref{CV_along_iso_LJ} and \ref{CV_along_iso_GF}. It 
cannot be otherwise, but there is more information here compared to those
figures. For example the adiabats are closer to the straight lines (in the
double-log representation) expected for IPL
systems, while the $C_V$-contours have more non-trivial shapes. Furthermore
a small variation of $C_V$ along an adiabat could hide a relatively large
difference in slope between $C_V$-contours and adiabats (since
$C_V$ is typically a relatively slowly varying function).


\section{\label{fluc_CV_contours}Fluctuation formula for generating contours of $C_V$}


Apart from investigating the variation of $C_V$ along an adiabat, it is of
interest to identify the contours of $C_V$; the non-constancy of $C_V$ along
an adiabat is equivalent to the statement that the $C_V$ contours do not
coincide with the adiabats, although we can expect them to be close for
Roskilde liquids. In practice we identify $C_V$ contours using the 
interpolation procedure described above, but it is potentially useful from a
theoretical point of view to have a fluctuation formula for the slope of 
these curves. This we derive in this section.

Since the variation of $C_V$ along an adiabat
(Eq.~\ref{Deriv_CV_curve_const_S}) involves the difference between
two triple correlations $ \angleb{(\Delta U)^2\Delta(W-\gamma U)}$ (which
vanishes for perfect correlation); it is tempting to speculate that the ratio

\begin{equation}
\frac{\angleb{(\Delta U)^2\Delta W} } { \angleb{(\Delta U)^3} },
\end{equation}
which equals $\gamma$ for perfect correlation, 
gives the slope of curves of constant $C_V$. In fact, it is not quite so simple.
The total derivative of $C_V$ with respect to $\ln\rho$ along
an arbitrary slope $g$ in the $(\ln\rho$, $\ln T)$ plane is

\begin{equation}\label{d_CV_slope_g}
\left(\frac{\partial C_V}{\partial\ln\rho}\right)_{[g]} = 
\left( \frac{\partial C_V}{\partial\ln\rho}\right)_T + 
g\left(\frac{\partial C_V}{\partial\ln T}\right)_\rho.
\end{equation}
We need to calculate the partial derivatives with respect to $T$ and $\rho$. 
From appendix~\ref{Deriv_CV_exponent}:

\begin{equation} \label{d_Cv_d_log_T_1}
  \left( \frac{\partial C_V}{\partial\ln T}\right)_\rho =
  -2\beta^2\angleb{(\Delta U)^2} - 
  \beta^3\frac{\partial\angleb{U^2}}{\partial\beta} 
  -2\beta^3\angleb{U} \angleb{(\Delta U)^2}.
\end{equation}
From  Eqs.~(\ref{dA_d_beta}) and \eqref{Delta_U_sq} we have

\begin{align}
\frac{\partial \angleb{U^2}}{\partial\beta} &= -\angleb{\Delta U\Delta(U^2)} \\
&= -\angleb{\Delta U\left(2\angleb{U}\Delta U + (\Delta U)^2 -\angleb{(\Delta U)^2}\right)} \\
&= -2\angleb{U}\angleb{(\Delta U)^2} - \angleb{(\Delta U)^3} 
\end{align}
Inserting this into Eq.~(\ref{d_Cv_d_log_T_1}) gives

\begin{align}
\left(\frac{\partial C_V}{\partial\ln T}\right)_\rho &= 
-2\beta^2\angleb{(\Delta U)^2} + \beta^3 \angleb{(\Delta U)^3} 
\end{align}
It might seem surprising that the third moment appears, since one expects
the limit of large $N$ that the distribution converges to a Gaussian, in
accordance with the central limit theorem. A closer look at the proof of that
theorem shows that when considering the summed variable (here the total
potential energy), all the so-called
cumulants are proportional to $N$, and both the second and third moments are
equal to the corresponding cumulants, and therefore proportional to $N$. It is
when one considers the average instead of the sum (potential 
energy per particle instead
of total potential energy) that one
finds the third moment and cumulant vanishing faster than the second ($1/N^2$ 
as opposed to $1/N$) in the limit of large $N$.

The density derivative of $C_V$

\begin{equation}
\left(\frac{\partial C_V}{\partial\ln\rho}\right)_T =
 \beta^2 \left(\frac{\partial\angleb{U^2}}{\partial\ln\rho}\right)_T -
\beta^2 2\angleb{U} \left(\frac{\partial\angleb{U}}{\partial\ln\rho}\right)_T 
\end{equation}
is evaluated in Appendix~\ref{Deriv_CV_exponent} with the result

\begin{equation}\label{CV_Log_Deriv}
\left(\frac{\partial C_V}{\partial\ln\rho}\right)_T =
-\beta^3\angleb{\Delta W(\Delta U)^2} +2\beta^2\angleb{\Delta U\Delta W}.
\end{equation}
The derivative of $C_V$ along an arbitrary slope $g$ is then

\begin{equation}\label{d_Cv_d_rho_slope_g}
\left(\frac{d C_V}{d\ln\rho}\right)_{[g]} = C_V 
\left(\frac{\beta\angleb{(\Delta U)^3}g - 
\beta\angleb{(\Delta U)^2\Delta W}}{\angleb{(\Delta U)^2}} + 2(\gamma-g) \right)
\end{equation}
Note that with $g=\gamma$ we recover Eq.~(\ref{Deriv_CV_curve_const_S}).
When the correlation is not perfect we can set this expression to
zero and solve for the slope $g$ which gives curves of constant $C_V$, 
now calling it $\gamma_{C_V}\equiv (\partial\ln T / \partial\ln\rho)_{C_V}$:

\begin{equation}
\gamma_{C_V} 
\left( \frac{\beta\angleb{(\Delta U)^3}}{\angleb{(\Delta U)^2}}-2\right)
=\frac{\beta\angleb{(\Delta U)^2\Delta W}}{\angleb{(\Delta U)^2}} -2\gamma
\end{equation}
or

\begin{equation}\label{gamma_CV_formula}
\gamma_{C_V} = \frac{\angleb{(\Delta U)^2\Delta W}-2T\gamma\angleb{(\Delta U)^2}}
{\angleb{(\Delta U)^3}-2T\angleb{(\Delta U)^2}} 
\end{equation}
Again we check the case of perfect correlation where we can replace $\Delta W$
by $\gamma\Delta U$ and see that we get $\gamma$ as we should. We can also write
 this as $\gamma$ plus a correction term:

\begin{equation}\label{gamma_CV_formula2}
\gamma_{C_V} = \gamma +
\frac{\angleb{(\Delta U)^2\epsilon}}
{\angleb{(\Delta U)^3}-2T\angleb{(\Delta U)^2}}
\end{equation}

\begin{figure}
\epsfig{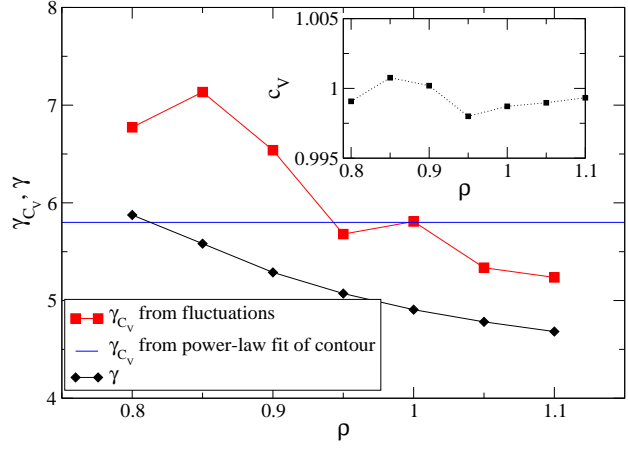}
\caption{\label{CV_exponent_along_CV_contour} Plot of $\gamma_{C_V}=(\partial
\ln T/\partial\ln\rho)_{C_V}$ estimated from
fluctuations, along the $C_V=1.0$ contour for the LJ system.
The contour was determined by interpolation. The horizontal line indicates the
slope found by fitting the contour to a power-law form for comparison. The 
decrease of $\gamma_{C_V}$ towards large densities is expected, just 
as with $\gamma$ (also shown) since at high densities we expect both to converge
to one third of the repulsive exponent, i.e., 4. The inset 
shows $C_V$ versus $\rho$
along the contour as a check that the contour was correctly determined.}
\end{figure}


\begin{figure}
\epsfig{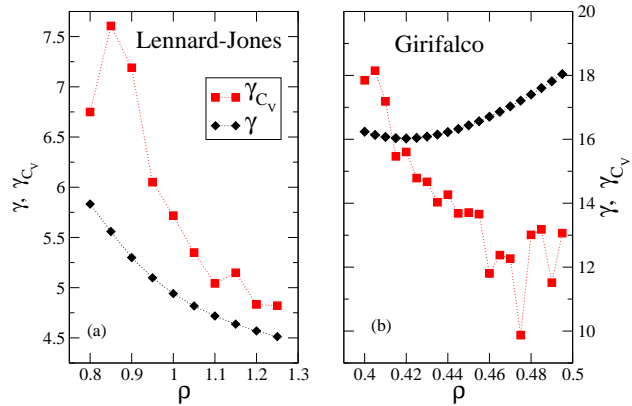}
\caption{\label{CV_exponent_along_adiabat} Plot of 
$\gamma_{C_V}=(\partial\ln T/\partial\ln\rho)_{C_V}$ estimated from
fluctuations, along (a) the adiabat including $\rho=0.8, T=0.8$ for
the LJ system and (b) the adiabat including $\rho=0.4, T=4.0$
for the Girifalco system, as functions of $\rho$, compared to $\gamma$.}
\end{figure}

Figure~\ref{CV_exponent_along_CV_contour} shows the fluctuation-determined
slope $\gamma_{C_V}$ of a $C_V$ contour in the 
$(\ln\rho, \ln T)$-plane along the $C_V=1.0$ contour of the LJ
system. We present the $C_V$-contour here to be able to check the validity
of the exponent: The (fixed) exponent determined by a fit of the contour to a 
power law is also indicated for comparison.
A clear trend is observed with $\gamma_{C_V}$ higher than $\gamma$, and
like the latter decreasing towards 4 as the density increases.
There is some scatter
due to the difficulty in determining third moments (compare 
the data for $\gamma$ which are based on second moments), so this would not be a
practical method for determining the contours. On the other hand,
if we are interested in knowing roughly how big the difference in slope
between an adiabat and a $C_V$-contour is, we do not need to simulate a 
$C_V$-contour--we can simulate a few state points, perhaps on an isochore, and
estimate the $\gamma_{C_V}$ from fluctuations. The scatter is not a big problem
if we are not using $\gamma_{C_V}$ to determine where to simulate next.
Fig.~\ref{CV_exponent_along_adiabat} compares $\gamma_{C_V}$ with $\gamma$
for both LJ and Girifalco system along an adiabat, and the trends are very
clear: the $C_V$-contours have definitely larger slope for the LJ system, closer
to 6 than 5 (they must converge to 4 at high density). For the 
Girifalco system the differences are quite dramatic, more so than the direct 
comparison of the contours in Fig.~\ref{const_S_vs_const_CV_GF} (where a 
logarithmic temperature scale was used). It is worth noting that all the data
here correspond to state points with $R>0.985$, i.e., very strong $U,W$
correlation, and that nothing special happens when the exponents are equal
 (e.g. $\rho\sim0.42$ in Fig.~\ref{CV_exponent_along_adiabat}(b)) (in any 
system one can define phase-space curves along which $\gamma-\gamma_{C_V} = 0$;
it would be significant only if a two-dimensional region of equality existed).

\section{Discussion}

\subsection{Roskilde liquids are more than, and more interesting than, IPL liquids}

IPL liquids are perfectly correlating and have perfect isomorphs---straight 
lines in the $(\ln\rho-\ln T)$ plane with slope given by one third of the IPL 
exponent. In this case the phase diagram is completely degenerate---the 
isomorphs are contours of excess entropy, $C_V$ and all structural and dynamical
properties (when expressed in reduced units). Liquids which have strong, but 
not perfect $U,W$ correlation are much more interesting: we can still identify
excellent isomorphs via Eq.~\eqref{adiabats_from_gamma}, as adiabats, but 
these are no longer constrained to be power laws; the effective exponent can
vary along an isomorph/adiabat and can exhibit non-trivial density 
dependence.\cite{Boehling/others:2013b} Moreover $C_V$ contours 
deviate now from the isomorphs/adiabats in a manner 
connected to the density dependence of $\gamma$.

It is interesting to compare the insight obtained from statistical 
mechanical versus thermodynamic considerations. Using statistical mechanics
---the arguments leading to Eq.~\eqref{gamma_derivatives_conjecture}---we 
have shown that $(\partial\gamma/ \partial T)_\rho$ vanishes when 
correlation is perfect, and this occurs only for (extended) IPL 
systems (see Eq.~\ref{extendedIPL}).
We have also argued that in liquids with strong but not perfect $U,W$ 
correlations the temperature
derivative is relatively small, therefore as a first approximation it can be
ignored, leaving the density dependence of $\gamma$ as a new characteristic for
a Roskilde
liquid. On the other hand the purely thermodynamic arguments
presented in Ref.~\onlinecite{Ingebrigtsen/others:2012} constrain only 
$(\partial\gamma/ \partial T)_\rho$ to be zero, leaving $\gamma$ free to depend
on density, which allows for the richer set of behaviors just mentioned. 
The thermodynamic argument leads more directly (and elegantly) to the empirical 
truth---that in practice $\gamma$'s temperature dependence is small compared
to its density dependence---while the statistical mechanical arguments 
fill in the details of why this is the case.

\subsection{Status of $n^{(2)}$ and relation between
different $\gamma$ derivatives}

The claim \eqref{gamma_derivatives_conjecture} needs to be 
thoroughly investigated by simulation for a wider range of systems as does the
validity of Eq.~\eqref{gamma_rho_estimate} as an estimate of $\gamma$. While we 
have argued these for high temperatures and densities, their validity could turn
out to depend on how strong $U,W$-correlation a liquid has, though it seems 
that $R>0.9$ is not necessarily required, that is, they apply
more generally than strong $U,W$ correlation.
One could imagine that it would be useful to derive
a fluctuation formula for $(\partial\gamma/\partial\rho)_S$. We have indeed
derived such a formula, see Appendix~\ref{gamma_deriv}, but it is not 
particularly simple, and we have not been able to use it to 
make a more rigorous theoretical 
connection with $(\partial\gamma/\partial T)_\rho$---even the sign is far from
obvious due to near cancellation of the various terms. 
Its usefulness in simulations
is also expected to be limited since it involves fluctuations
of the so-called hypervirial (the quantity used to determine the bulk modulus
from fluctuations\cite{Allen/Tildesley:1987}) which is not typically available 
in an MD simulation. On the other hand, given our
results, one can use the quantity $\angleb{(\Delta U)^2\epsilon}$ or the
formula for $\gamma_{C_V}$ to determine the sign of
 $(\partial\gamma/\partial\rho)_S$ from a simulation of a single state point.


\subsection{Adiabats versus $C_V$ contours in non-Roskilde-simple 
liquids}

\begin{figure}
\epsfig{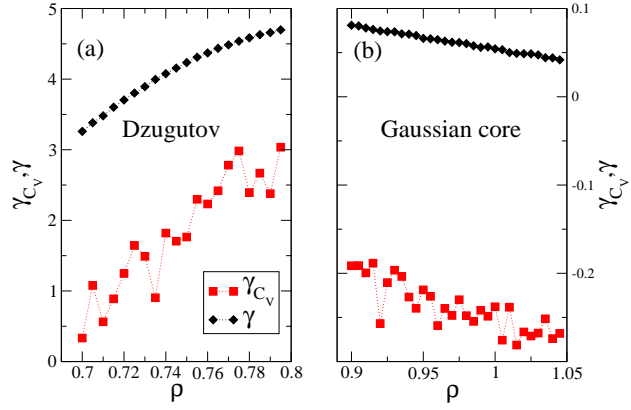}
\caption{\label{CV_exponent_along_adiabat_DZ_GAU} Plot of 
$\gamma_{C_V}=(\partial\ln T/\partial\ln\rho)_{C_V}$ estimated from
fluctuations, for (a) the Dzugutov system along the adiabat 
including $\rho=0.70, T=0.70$ and (b) the Gaussian core system along the 
adiabat starting at $\rho=0.90, T=0.75$,
as functions of $\rho$, compared to $\gamma$.}
\end{figure}

It is interesting to consider a non-simple liquid, where there
is no reason to expect that $C_V$-contours at all coincide with adiabats 
(i.e. there are not good isomorphs). We have done so for two liquids
without actually determining the $C_V$-contours; instead we just calculated the
exponent $\gamma_{C_V}$ from the fluctuations. As mentioned above this is 
accurate enough to
give an idea of the trends, in particular which way the $C_V$-contours are
oriented with respect to the adiabats. The first example is the Dzugutov 
fluid.\cite{Dzugutov:1992} Fig.~\ref{CV_exponent_along_adiabat_DZ_GAU} 
shows $\gamma_{C_V}$ and $\gamma$ for this system along an adiabat. In the range
shown $R$ takes values from $\sim 0.56$ to $\sim0.84$. As the figure shows
$\gamma_{C_V}$ is substantially smaller than $\gamma$. We can note also that
this is consistent with the positive slope $d\gamma/d\rho$, and suggests the 
arguments leading to Eq.~\eqref{gamma_derivatives_conjecture} do not necessarily
require strong $W,U$ correlation. Another example
is the Gaussian core potential,\cite{Stillinger:1976} for which data is also 
shown in 
Fig.~\ref{CV_exponent_along_adiabat_DZ_GAU}. In this case there is almost 
no $W,U$ correlation; $0.16>R>0.06$, and in fact $\gamma_{C_V}$ and $\gamma$ even
have opposite sign (although both are close to zero). Moreover this system 
clearly violates Eq.~\eqref{gamma_derivatives_conjecture},
since $\gamma$ decreases with density on the adiabat shown, which should 
correspond to the case $\gamma_{C_V}>\gamma$ (as in the LJ case); this is not
surprising since it does not have a hard core.

\subsection{Relevance of adiabats versus $C_V$ contours}

\begin{figure}
\epsfig{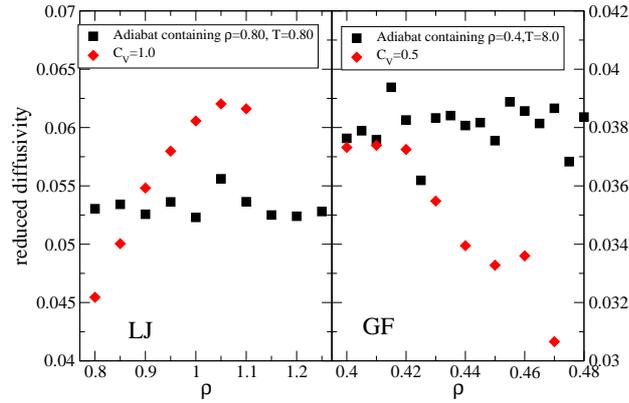}
\caption{\label{CompareD_Isomorph_CV_Contour} Diffusivity in reduced units 
versus density along an adiabat and along a $C_V$-contour 
for the LJ and GF systems. It is more or less invariant on the adiabats but not
on the $C_V$-contours.}
\end{figure}

In our simulation studies of isomorphs, the procedure has 
always been to use Eq.~\eqref{adiabats_from_gamma} to generate adiabats 
(straightforward, since an accurate estimate of $\gamma$ is readily computed 
from the $W,U$ fluctuations)
and then examine to what extent the other isomorph-invariant quantities are
actually invariant along these curves. One could also
generate $C_V$ contours and check for invariance along them. While it is not 
obvious that adiabats are more fundamental, Rosenfeld has proposed that 
transport properties are in fact
governed by the excess entropy.\cite{Rosenfeld:1999} Given the not
insignificant difference between adiabats and $C_V$-contours it is worth 
checking explicitly whether measures of dynamics are more invariant along one
versus the other. This is done in Fig.~\ref{CompareD_Isomorph_CV_Contour} for
the reduced diffusivity $\tilde D\equiv\left( \rho^{1/3}\sqrt{m/T}\right) D$.
It is clear that by the this measure, the dynamics are more invariant along
adiabats than along $C_V$-contours, consistent with Rosenfeld's theory. 
We note also that the adiabats seem to be simpler than the $C_V$-contours in 
that the exponent $\gamma$ varies less than the exponent $\gamma_{C_V}$. This is
true for the all the systems presented here including simple and non-simple
ones. This implies $\gamma$ is more practical as a liquid
characteristic than $\gamma_{C_V}$ and suggests that adiabats provide a
more useful, and fundamental basis for describing the phase diagram 
than $C_V$-contours. In fact a $(\rho,S)$ phase diagram would be 
consistent with the traditional
starting point of statistical mechanics---a function
$U(S,V)$ expressing the dependence of internal energy on entropy and volume 
(though typically the total entropy, not $S$, is considered).

\section{Conclusion}


We have derived several exact results relating to Roskilde-simple liquids
(previously termed strongly correlating liquids)
in the form of fluctuation formulas
for various thermodynamic derivatives. These include the derivative (with 
respect to $\ln\rho$) of an arbitrary NVT averaged dynamical variable 
along a configurational adiabat, Eq.~\eqref{deriv_A_const_S}, the derivative of
$C_V$ along an adiabat, Eq.~\eqref{Deriv_CV_curve_const_S}, the temperature
derivative of $\gamma$ itself on an isochore, 
Eq.~\eqref{d_C_V_d_rho_S_d_gamma_dT}, and the slope of contours of $C_V$ in the 
$(\ln\rho, \ln T)$ plane, Eq.~\eqref{gamma_CV_formula}.
In addition to the exact formulas we have argued that when 
$d\gamma/d\rho$ is negative (positive) one expects that
 $(\partial C_V/\partial\rho)_S$ is
 positive (negative) and that the slopes of $C_V$-contours are greater (less)
than those of adiabats. This we have tested with two model Roskilde-simple 
liquids, the 
Lennard-Jones fluid with $d\gamma/d\rho<0$ and the Girifalco potential which
has $d\gamma/d\rho<0$ at low density but switches to $d\gamma/d\rho>0$ at
high density. From this argument emerged a 
claim, Eq.~\eqref{gamma_derivatives_conjecture} equating the sign of the
temperature derivative of $\gamma$ to the density derivative along an adiabat
for a wide class of liquids (wider than Roskilde-simple liquids).
Finally we note that the data presented here provide support for the use of
the $n^{(2)}$ exponent, determined purely by the pair potential, as a quick and
convenient way to estimate $\gamma$ and its density dependence.
\appendix

\section{\label{dA_d_ln_rho_deriv}Derivation of Eq.~\eqref{dA_d_ln_rho}}

As in appendix A of Ref.~\onlinecite{Bailey/others:2008b}, we use a 
discrete-state notation for convenience, such that $A_i$ is the value of 
observable $A$ in microstate $i$ and the (configurational) partition function is
$Z=\sum_i\exp(-\beta U_i)$. We have

\begin{align}
\left(\frac{\partial\angleb{A}}{\partial\ln\rho}\right)_T &= 
\frac{1}{Z} \frac{\partial\sum_i A_i \exp(-\beta U_i)}{\partial\ln\rho} 
- \frac{1}{Z^2} \sum_iA_i\exp(-\beta U_i)\frac{\partial\sum_j \exp(-\beta U_j)}{\partial\ln\rho} \\ 
&= \frac{1}{Z} \sum_i \left(\frac{\partial A_i}{\partial\ln\rho} \exp(-\beta U_i) + A_i \exp(-\beta U_i)(-\beta)\frac{\partial U_i}{\partial\ln\rho} \right)
-\frac{\sum_i A_i \exp(-\beta U_i)}{Z^2}\sum_j\exp(-\beta U_j)(-\beta)\frac{\partial U_i}{\partial\ln\rho} \\
&= \angleb{\frac{\partial A}{\partial\ln\rho}} - \beta\left(\angleb{AW} - \angleb{A}\angleb{W}\right) \\
&= \angleb{\frac{\partial A}{\partial\ln\rho}} - \beta\angleb{\Delta A\Delta W}.
\end{align}
In the second last step the definition of the virial for a micro-configuration,
$W_i\equiv (\partial U_i/\partial\ln\rho)$ was used; the density 
derivative is understood to mean that the reduced coordinates are held fixed
while the volume is changed.

\section{\label{Derivation_C_V_deriv}Derivation of Eq.~\eqref{Deriv_CV_curve_const_S}}

Here we give the details of the derivation of the expression for the derivative
of $C_V$ at constant $S$. Writing the variance of $U$ as
$\angleb{(\Delta U)^2}= \angleb{U^2}-\angleb{U}^2$
allows us to use Eq.~(\ref{deriv_A_const_S}) to take the 
derivative of $\angleb{U^2}$  
and Eq.~(\ref{dU_d_rho_S}) to differentiate $\angleb{U}$. 

\begin{align}
 \left(\frac{\partial\angleb{(\Delta U)^2} }{\partial
\ln\rho}\right)_S &= \angleb{\frac{\partial U^2}{\partial  \ln\rho}} -\beta \angleb{\Delta(U^2)\Delta(W-\gamma U)} - 2\angleb{U}\angleb{W}\\
&=\angleb{2UW} - \beta \angleb{\Delta(U^2)\Delta(W-\gamma U)} - 2\angleb{U}\angleb{W} \\
&= 2\angleb{\Delta U\Delta W} - \beta \angleb{\Delta(U^2)\Delta(W-\gamma U)} \\
&= 2\gamma\angleb{(\Delta U)^2} - \beta \angleb{\Delta(U^2)\Delta(W-\gamma U)}
\end{align}
where we have used Eqs.~(\ref{gamma_definition}) to write the covariance
$\angleb{\Delta U \Delta W}$ in terms of the variance of $U$.
Inserting this result with Eq.~(\ref{d_Cv_d_rho_S_2}) gives
the relatively simple formula

\begin{equation}
\left( \frac{\partial C_V}{\partial \ln\rho}\right)_S =
 - \beta^3 \angleb{\Delta(U^2)\Delta(W-\gamma U)} = 
 - \beta^3 \angleb{\Delta(U^2)\epsilon} 
\end{equation}
We make one more change by writing $U=\angleb{U}+\Delta U$, so that

\begin{align}
\Delta(U^2) &= U^2 - \angleb{U^2} \\
&=\angleb{U}^2 + 2 \angleb{U} \Delta U + (\Delta U)^2 - (\angleb{U}^2  +
\angleb{(\Delta U)^2}) \\
&=  2 \angleb{U} \Delta U +  (\Delta U)^2 - \angleb{(\Delta U)^2}\label{Delta_U_sq}
\end{align}
When this is correlated with $\epsilon=\Delta W - \gamma \Delta U$, the first 
term vanishes because of Eq.~(\ref{corr_DeltaU_epsilon}) and the last term
vanishes because $\angleb{\epsilon}=0$. Thus 
$\angleb{\Delta(U^2)\epsilon}=\angleb{(\Delta U)^2\epsilon}$ and we arrive at
 Eq.~(\ref{Deriv_CV_curve_const_S}).

\section{\label{generating_adiabats}Generating configurational adiabats}

Eq.~(\ref{adiabats_from_gamma}) indicates a general procedure for generating
adiabats: (1) evaluate $\gamma$ from the fluctuations at the 
current state point; (2) choose a small change in density, say of order 1\% or 
less; (3) use Eq.~(\ref{adiabats_from_gamma}) to determine the corresponding 
change in temperature:

\begin{align}
\rho_{n+1} &= \rho_n + \delta\rho \\
T_{n+1} &= T_n \left(\rho_{n+1}/\rho_n\right)^{\gamma_n}
\end{align}

We have used this method for the Girifalco system with $\delta\rho=0.005$ for
values of $\rho$ between 0.4 and 0.5. For 
generalized Lennard-Jones systems there is now an analytic expression for the
$\rho$-dependence of $\gamma$ which allows large changes in $\rho$, the 
so-called ``long jump method'':\cite{Ingebrigtsen/others:2012, 
Boehling/others:2013a}

\begin{align}\label{long_jump_formula}
\rho_{n+1} &= \rho_n + \delta\rho \\
T_{n+1} &= T_n h(\alpha_n, \rho_{n+1})/h(\alpha_n, \rho_n)
\end{align}
where the energy/temperature scaling function $h(\alpha, \rho)$ is defined by
(see Refs.~\onlinecite{Ingebrigtsen/others:2012, Boehling/others:2013a}; the 
normalization is such that $h(\alpha,1)=1$).

\begin{equation}\label{h_of_rho}
h(\alpha,\rho) = \alpha\rho^4 + (1-\alpha)\rho^2
\end{equation}
Here $\alpha$ is a parameter which according to the theory of 
isomorph---i.e., assuming perfect isomorphs for LJ systems---is a constant.
More generally one may expect that it is fixed for a given isomorph, but can 
vary weakly
among isomorphs, analogous\footnote{There is in fact a close connection between
$h(\rho)$ and $n^{(2)}$; by identifying Eq.~\ref{gamma_rho_estimate} with the 
logarithmic derivative of $h(\rho)$ we find that $h(\rho)$ can be expressed in
terms of the curvature of the pair potential, and moreover it becomes clear how
to include dependence on $S$ in $h(\rho)$. This connection will be discussed in
more detail elsewhere.\cite{Boehling/others:2013b}} to $\Lambda(S)$ in 
Eq.~\eqref{gamma_rho_estimate}. It can be evaluated at a given density via
(since $\gamma=d\ln(h)/d\ln(\rho)$\cite{Ingebrigtsen/others:2012})

\begin{equation}
\alpha = (\gamma-2) / (4\rho^2-2-\gamma\rho^2+\gamma)
\end{equation}
(at $\rho=1$ this becomes simply $\gamma/2-1$). 
Since the theory is not exact, and $\alpha$ determined this way will also 
vary weakly along the isomorph, in
order to get the best determination of the adiabats we 
re-evaluate $\alpha$ at each state point. It therefore also has an index $n$. 
We observe a systematic variation in $\alpha$ of at most 0.5\% for a given 
adiabat, and a few percent variation 
between adiabats. We have used the long-jump formula
for the LJ system with $\delta\rho=0.05$ for values of 
$\rho$ between 0.8 and 1.4.
We noticed more noise in the data for the Girifalco system, but have not checked
whether this is due to not having a long-jump formula or to differences in
effective sampling rate (because of different relaxation times) 
giving different statistical errors.

\section{\label{Deriv_CV_exponent}Derivation of $C_V$ exponent}

The temperature derivative of $C_V$, 
Eq.~(\ref{d_Cv_d_log_T_1}), is obtained as follows:

\begin{align}
\left(\frac{\partial C_V}{\partial\ln T}\right)_\rho 
&= -\frac{\partial}{\partial \ln \beta}
\left(\beta^2\angleb{(\Delta U)^2}\right) =
 -\beta \frac{\partial}{\partial \beta}
\left(\beta^2\angleb{(\Delta U)^2}\right) \\
&= -2\beta^2\angleb{(\Delta U)^2} -\beta^3\frac{\partial}{\partial\beta} 
\left( \angleb{U^2}-\angleb{U}^2\right) \\
&= -2\beta^2\angleb{(\Delta U)^2} - 
\beta^3\frac{\partial\angleb{U^2}}{\partial\beta} 
+2\beta^2\angleb{U} \beta\frac{\partial \angleb{U}}{\partial\beta} \\
&= -2\beta^2\angleb{(\Delta U)^2} - 
\beta^3\frac{\partial\angleb{U^2}}{\partial\beta} 
-2\beta^2\angleb{U} T C_V \\
&= -2\beta^2\angleb{(\Delta U)^2} +
\beta^3\angleb{(\Delta (U^2) \Delta U}
-2\beta^3\angleb{U} \angleb{(\Delta U)^2}.
\end{align}
In the last line Eq.~\eqref{dA_d_ln_T} was used. We can simplify by using 
Eq.~\eqref{Delta_U_sq}:

\begin{align}
\left(\frac{\partial C_V}{\partial\ln T}\right)_\rho 
&= -2\beta^2\angleb{(\Delta U)^2} +
\beta^3\left(2\angleb{U}\angleb{(\Delta U)^2} +\angleb{(\Delta U)^3}\right)
-2\beta^3\angleb{U} \angleb{(\Delta U)^2}\\
&= -2\beta^2\angleb{(\Delta U)^2} + \beta^3 \angleb{(\Delta U)^3}.
\end{align}

For the density derivative of $C_V$ we have likewise

\begin{equation}
\left(\frac{\partial C_V}{\partial\ln\rho}\right)_T =
 \beta^2 \left(\frac{\partial\angleb{U^2}}{\partial\ln\rho}\right)_T -
\beta^2 2\angleb{U} \left(\frac{\partial\angleb{U}}{\partial\ln\rho}\right)_T
\end{equation}
Starting with the second term, using Eq.~\eqref{dA_d_ln_rho}

\begin{equation}
 \left(\frac{\partial\angleb{U}}{\partial\ln\rho}\right)_T = 
-\beta\angleb{\Delta W\Delta U} + \angleb{W}
\end{equation}
while the first gives, also using Eqs.~\eqref{dA_d_ln_rho} and 
\eqref{Delta_U_sq}

\begin{align}
 \left(\frac{\partial\angleb{U^2}}{\partial\ln\rho}\right)_T &= 
-\beta\angleb{\Delta W\Delta (U^2)} + \angleb{2UW} \\
&=  -2\beta\angleb{U}\angleb{\Delta W \Delta U)} -
\beta\angleb{\Delta W(\Delta U)^2)} + 2\angleb{UW}
\end{align}
Combining the two terms then gives

\begin{align}
\left(\frac{\partial C_V}{\partial\ln\rho}\right)_T &= 
-\beta^3 2\angleb{U}\angleb{\Delta U\Delta W}
-\beta^3\angleb{\Delta W(\Delta U)^2} +2\beta^2\angleb{UW}
+2\angleb{U}\beta^3 \angleb{\Delta W\Delta U}
-2\angleb{U} \beta^2 \angleb{W} \\
&= -\beta^3\angleb{\Delta W(\Delta U)^2} +2\beta^2\angleb{\Delta U\Delta W}
\end{align}
which is Eq.~(\ref{CV_Log_Deriv}). 
Now we can assemble the derivative of $C_V$ along an arbitrary slope $g$ 
(Eq.~\eqref{d_CV_slope_g}):

\begin{align}
\left(\frac{d C_V}{d\ln\rho}\right)_{[g]} &= 
-\beta^3\angleb{\Delta W(\Delta U)^2} +2\beta^2\angleb{\Delta U\Delta W} 
+g\left(-2\beta^2\angleb{(\Delta U)^2} + \beta^3 \angleb{(\Delta U)^3} \right)\\
&= \beta^2\angleb{(\Delta U)^2} \left(
-\beta\angleb{\Delta W(\Delta U)^2} / \angleb{(\Delta U)^2}+2\gamma+ 
g\left( -2 + \beta \angleb{(\Delta U)^3}/ \angleb{(\Delta U)^2} \right)
\right)
\end{align}
which can be rewritten as Eq.~(\ref{d_Cv_d_rho_slope_g}).

\section{\label{gamma_deriv}Fluctuation formula for the derivative of $\gamma$}

We include here, omitting the derivation, the fluctuation formula for the
derivative of $\gamma$ with respect to $\ln\rho$ at constant $S$. The quantity
$X\equiv dW/d\ln\rho$ is the hypervirial, which appears in fluctuation
expressions for the bulk modulus.\cite{Allen/Tildesley:1987}

\begin{equation}\label{d_gamma_d_rho_S_Dvar_Dtriple}
\left(\frac{\partial \gamma}{\partial \ln\rho} \right)_S
= \frac{1}{\angleb{(\Delta U)^2}} \left(
\angleb{(\Delta W)^2} + \angleb{\Delta U\Delta X}
- 2\gamma^2\angleb{(\Delta U)^2} 
-\beta\angleb{\Delta U\epsilon^2}
\right)
\end{equation}
For IPL systems we have $\Delta X=\gamma\Delta W$ and $\epsilon\equiv0$, so that
the derivative is zero.

\acknowledgments

The centre for viscous liquid dynamics ``Glass and Time'' is sponsored by the 
Danish National Research Foundation's grant DNRF61.


\end{document}